\documentclass[12pt]{article}
\usepackage{amsmath}
\usepackage{graphicx}
\usepackage{amsfonts}
\usepackage{amssymb}

\begin{document}

\title{Discontinuity of Electrostatic Fields Across a Planar Surface With Multipolar
Surface Charge Density}
\author{S. Gov, S. Shtrikman\thanks{Also with the Department of Physics, University of
California, San Diego, La Jolla, 92093 CA, USA.} and H. Matzner\\Department of Electronics, Weizmann Institute of Science, \\Rehovot 76100, Israel.}
\date{}
\maketitle
\begin{abstract}
We use covariance and dimensional analysis to find expressions for the
discontinuity of the potential and normal electric field across a flat surface
with\emph{ multipolar} charge surface density in vacuum. In particular we show
that
\[
\delta E_{z}=\sum\limits_{l=0}^{\infty}\sum\limits_{m=-l}^{l}a_{l,m}%
(\partial_{+})^{\frac{1}{2}(l+m)}(\partial_{-})^{\frac{1}{2}(l-m)}q_{l,m}(x,y)
\]
and that
\[
\delta\Psi=\sum\limits_{l=0}^{\infty}\sum\limits_{m=-l}^{l}b_{l,m}%
(\partial_{+})^{\frac{1}{2}[(l-1)+m]}(\partial_{-})^{\frac{1}{2}%
[(l-1)-m]}q_{l,m}(x,y)
\]
Here, $E_{z}$ is the normal electric field, $\Psi$ is the electrostatic
potential, $q_{l,m}(x,y)$ is the surface density of the ($l^{\text{th}%
},m^{\text{th}}$) electric multipole and $\partial_{\pm}\equiv\partial
/\partial x\pm i\partial/\partial y$. The prefactors $a_{l,m}$ and $b_{l,m}$
in these relations are calculated by explicitly evaluating the field of a
localized unit multipole, both above and below the surface.
\end{abstract}

\section{Introduction}

One of the common problems in electrostatics is the determination of electric
field or potential in the presence of a surface distribution of charges. The
knowledge of the discontinuity in the electric field across charged surfaces
is useful in solving boundary value problems in electrostatic. Gauss's law
allows us to write down a result directly. According to Gauss law the
discontinuity of the normal electric field across a surface charge density
$\sigma$ is $4\pi\sigma$ (in CGS units)\cite{jump1}. Another important result
is the discontinuity of the electrostatic potential across a dipolar layer
with the dipoles pointing normal to the surface. In this case\cite{jump2}
$\delta\Psi=4\pi P_{z}$ where $\Psi$ is the potential and $P_{z}$ is the
dipole surface density.

In this paper we study the electric field produced by a flat surface, in
vacuum, having general ($l^{th},m^{th}$) multipole surface density and give
expressions for the discontinuity in the normal and tangential electric field.
We show that the discontinuity in the normal electric field is given by
\begin{align}
\delta E_{z}(x,y)  &  =%
{\displaystyle\sum\limits_{l=0}^{\infty}}
{\displaystyle\sum\limits_{m=-l}^{l}}
2\pi\left[  1+(-1)^{l+m}\right]  \sqrt{\frac{4\pi}{2l+1}}\frac{1}%
{\sqrt{(l+m)!(l-m)!}}\times\label{eq0.1}\\
&  \ \ (\partial_{+})^{\frac{1}{2}(l+m)}(\partial_{-})^{\frac{1}{2}%
(l-m)}q_{l,m}(x,y)\nonumber
\end{align}
whereas the discontinuity in the potential is and
\begin{align}
\delta\Psi(x,y)  &  =%
{\displaystyle\sum\limits_{l=0}^{\infty}}
{\displaystyle\sum\limits_{m=-l}^{l}}
2\pi\left[  1-(-1)^{l+m}\right]  \sqrt{\frac{4\pi}{2l+1}}\frac{1}%
{\sqrt{(l+m)!(l-m)!}}\times\label{eq0.3}\\
&  \ (\partial_{+})^{\frac{1}{2}[(l-1)+m]}(\partial_{-})^{\frac{1}%
{2}[(l-1)-m]}q_{l,m}(x,y)\nonumber
\end{align}
Here,

\begin{itemize}
\item $q_{l,m}(x,y)$ is the $(l^{th},m^{th})$ multipolar surface charge
density over the plane.

\item $\partial_{\pm}\equiv\partial/\partial x\pm i\partial/\partial y$

\item $E_{z}$ is the normal electric field

\item $\Psi$ the potential.
\end{itemize}

From another point of view Eq.(\ref{eq0.1}) may be interpreted as the
discontinuity in the normal electric field due to an `effective' surface
charge density that is given by
\begin{align*}
\sigma^{eff}(x,y)  &  =%
{\displaystyle\sum\limits_{l=0}^{\infty}}
{\displaystyle\sum\limits_{m=-l}^{l}}
\frac12\left[  1+(-1)^{l+m}\right]  \sqrt{\frac{4\pi}{2l+1}}\frac
1{\sqrt{(l+m)!(l-m)!}}\times\\
&  (\partial_{+})^{\frac12(l+m)}(\partial_{-})^{\frac12(l-m)}q_{l,m}(x,y)
\end{align*}
Similarly, the right hand side Eq.(\ref{eq0.3}) is equivalent to an
`effective' surface dipolar layer given by
\begin{align*}
P_{z}^{eff}(x,y)  &  =%
{\displaystyle\sum\limits_{l=0}^{\infty}}
{\displaystyle\sum\limits_{m=-l}^{l}}
2\pi\left[  1-(-1)^{l+m}\right]  \sqrt{\frac{4\pi}{2l+1}}\frac1{\sqrt
{(l+m)!(l-m)!}}\times\\
&  \ (\partial_{+})^{\frac12[(l-1)+m]}(\partial_{-})^{\frac12[(l-1)-m]}%
q_{l,m}(x,y)
\end{align*}

\section{Dimensional Analysis \& Covariance}

\subsection{Definition of the problem.}

Consider a surface at $z=0$ charged with $(l^{th},m^{th})$ multiples whose
surface density is $q_{l,m}(x,y)$. Thus, $q_{l,m}(x,y)\delta x\delta y$ is the
strength of the $(l^{th},m^{th})$ multipole located at the point $(x,y)$ on
the surface. If ,for example, $q_{l,m}(x,y)$ represents a unit multipolar
charge at the origin then $q_{l,m}(x,y)=\delta(x)\delta(y)$ producing a
potential given by
\[
\Psi=\frac{4\pi}{2l+1}\frac{Y_{l,m}(\theta,\varphi)}{r^{l+1}}%
\]
To fully characterize the surface charge all the multipole density functions
must be given (recall that $l=0,1,2,...$ and $m=-l,-(l-1),..,(l-1),l$) . For
example, for a surface with `regular' charge density, given by $\sigma(x,y)$,
the $q_{l,m}$'s are as follows: $q_{0,0}(x,y)=\sigma(x,y)/\sqrt{4\pi}$ and
$q_{l,m}(x,y)=0$ for higher multiples. These multipole moments give rise to an
electric field above and below the surface. The purpose of this paper is to
find expressions for the \emph{discontinuity} of the normal field and
potential in terms of $q_{l,m}(x,y)$. Thus, for instance, in the example given
above we would expect to find that $\delta E_{z}(x,y)\propto q_{00}%
(x,y)\propto\sigma(x,y)$ where $E_{z}$ is the normal component of the electric field.

\subsection{The permissible operators, their transformation rules and dimensions.}

In order to find expressions for the discontinuity of the field we use
covariance and dimensional analysis. We start by stating all the operations
that can be performed on $q_{l,m}(x,y)$ together with their transformation
rule under rotation and their dimensions. Using these we show that there is
only one possible form for the expression, up to an additional multiplicative
factor, for the discontinuity of the field which is both covariant and
dimensionally correct.

When the system is rotated by an angle $\alpha$ around the $\hat{z}$ axis we
note that $q_{l,m}$ transforms according to $q_{l,m}\rightarrow e^{im\alpha
}q_{l,m}$ (see Appendix A). The potential $\Psi$ is a scalar under rotation.
Next, any expression for the fields discontinuity must depend \emph{locally}
on $q_{l,m}(x,y)$ as the field near the surface (i.e. $\mathbf{E}%
(x,y,z\rightarrow0^{\pm})$ ) is determined mostly by the presence of charge
distribution at the vicinity of the point $(x,y,z=0)$.

The first operators that come to mind are multiplication by $x$ and%
$\backslash$%
or $y$. These operators, however, are not permissible since their use would
break the translational symmetry that is supposed to hold. The only operators
that are both local and translational invariant are $\partial/\partial x$ and
$\partial/\partial y$. However, each of these operators separately has no
definite transformation under rotation around the $\hat z$ axis. Operators
which \emph{do}\textbf{\ }have definite transformation rules are easily
constructed from these. Consider the operators $\partial_{\pm}\equiv$
$\partial/\partial x\pm i\partial/\partial y$ . When the coordinate system is
rotated around the $\hat z$ axis by an angle $\alpha$ to form a new coordinate
system- $(x^{^{\prime}},y^{^{\prime}})$ the newly defined operators transform
according to (see Appendix A) $\partial_{\pm}\rightarrow e^{\mp i\alpha
}\partial_{\pm}$. Note also that the operator $\partial_{-}\partial_{+}$ is a
scalar in this sense and is nothing but the two-dimensional Laplacian (which
can also be written as $\partial_{-}\partial_{+}$ since $\partial_{+}$ and
$\partial_{-}$ commutes).

We now turn to the transformation rules for the electric field. Obviously, the
component of the field along the $\hat z$ axis acts as a scalar under the
rotation. As for the tangential components we again construct new fields
defined by $E_{\pm}=E_{x}\pm iE_{y}$. As $E_{\pm}=-\partial_{\pm}\Psi$ and
$\Psi$ is a scalar the transformation rules for $E_{\pm}$ are the same as
those for $\partial_{\pm}$, namely $E_{\pm}\rightarrow e^{\mp i\alpha}E_{\pm}$.

The following table summarizes the transformation rules under rotation for
each of the quantities above together with their dimensions:%

\[%
\begin{tabular}
[c]{|c|c|}\hline
Transformation Rule & Dimensions\\\hline\hline
$q_{l,m}\rightarrow e^{im\alpha}q_{l,m}$ & $[Charge][Length]^{l-2}$\\\hline
$\partial_{+}\rightarrow e^{-i\alpha}\partial_{+}$ & $[Length]^{-1}$\\\hline
$\partial_{-}\rightarrow e^{+i\alpha}\partial_{-}$ & $[Length]^{-1}$\\\hline
$\Psi\rightarrow\Psi$ & $[Charge][Length]^{-1}$\\\hline
$E_{z}\rightarrow E_{z}$ & $[Charge][Length]^{-2}$\\\hline
$E_{+}\rightarrow e^{-i\alpha}E_{+}$ & $[Charge][Length]^{-2}$\\\hline
$E_{-}\rightarrow e^{+i\alpha}E_{-}$ & $[Charge][Length]^{-2}$\\\hline
$\partial_{+}\partial_{-}\rightarrow\partial_{+}\partial_{-}$ & $[Length]^{-2}%
$\\\hline
\end{tabular}
\]

\subsection{Covariant expression for the discontinuity in the normal electric
field $E_{z}$.}

We start with $q_{l,m}(x,y)$ which transforms like $e^{im\alpha}$. To find the
discontinuity in $E_{z}(x,y)$ we operate with $\partial_{+}$ a total number of
$M$ times and with $\partial_{-}$ a total number of $N$ times to get
\begin{equation}
\delta E_{z}(x,y)\propto(\partial_{+})^{M}(\partial_{-})^{N}q_{l,m}(x,y)
\label{eq.1}%
\end{equation}
Applying the transformation rules on both sides of the equation yields the
first connection between $N$ and $M$, namely:
\begin{equation}%
\begin{array}
[c]{c}%
1=e^{-i\alpha M}e^{+i\alpha N}e^{i\alpha m}\\
\Updownarrow\\
0=-M+N+m
\end{array}
\end{equation}
Dimensional analysis, on the other hand, requires that
\begin{equation}%
\begin{array}
[c]{c}%
\lbrack Charge][Length]^{-2}=[Charge][Length]^{-M-N+l-2}\\
\Updownarrow\\
-2=-M-N+l-2
\end{array}
\label{eq.3}%
\end{equation}
Solving for $M$ and $N$ gives
\begin{equation}%
\begin{array}
[c]{cc}%
M=\frac{1}{2}(l+m) & N=\frac{1}{2}(l-m)
\end{array}
\label{eq.4}%
\end{equation}
Thus, the general expression for the discontinuity in $E_{z}$ is
\begin{equation}
\delta E_{z}(x,y)=%
{\displaystyle\sum\limits_{l=0}^{\infty}}
{\displaystyle\sum\limits_{m=-l}^{l}}
a_{l,m}(\partial_{+})^{\frac{1}{2}(l+m)}(\partial_{-})^{\frac{1}{2}%
(l-m)}q_{l,m}(x,y) \label{eq.5}%
\end{equation}
where an additional multiplicative factor has been included to make an
equality out of the proportionality ( Note that the multiplicative factor
depends on $l$ and $m$ as there is no reason to assume otherwise). The result
is further summed over all the multiples .

\subsection{Covariant expressions for the discontinuity in the potential
$\Psi$ and the tangential electric field $E_{+}$ \& $E_{-}$.}

Similar arguments lead to the following result for the discontinuity in $\Psi
$:
\begin{equation}
\delta\Psi(x,y)=%
{\displaystyle\sum\limits_{l=0}^{\infty}}
{\displaystyle\sum\limits_{m=-l}^{l}}
b_{l,m}(\partial_{+})^{\frac{1}{2}[(l-1)+m]}(\partial_{-})^{\frac{1}%
{2}[(l-1)-m]}q_{l,m}(x,y) \label{eq.6}%
\end{equation}

\section{Determination of $a_{l,m}$ and $b_{l,m}$.}

In this section we actually resolve the problem of finding the discontinuity
of the fields across a multipolar layer. This time however, the exact
mathematical expressions are derived together with the required multiplicative factor.

\subsection{Fourier Decomposition of Spherical Harmonics Moments.}

Let $\Phi_{l,m}(r,\theta,\varphi)$ denote the potential due to a
\emph{localized} $(l,m)$-moment with unit strength. Thus \cite{jack1},
\begin{equation}
\Phi_{l,m}(r,\theta,\varphi)=\frac{4\pi}{2l+1}\frac{Y_{l,m}(\theta,\varphi
)}{r^{l+1}} \label{eq.1.0}%
\end{equation}
The total potential, $\Psi_{lm}(x,y,z)$, is then given by a superposition of
all the contributions from the surface, namely%

\begin{equation}
\Psi_{lm}(x,y,z)=%
{\displaystyle\int}
dx^{^{\prime}}\int dy^{^{\prime}}q_{lm}(x^{^{\prime}},y^{^{\prime}})\Phi
_{lm}(x-x^{^{\prime}},y-y^{^{\prime}},z)\equiv q_{lm}\ast\Phi_{lm}
\label{eq1.1}%
\end{equation}
where the notation $\ast$ stands for the \emph{two dimensional} convolution operation.

The unit multipole potential may be rewritten in terms of its Fourier
transform over the $(x,y)$ plane. Since $\Phi_{lm}$ is singular at the origin,
the plane wave expansion for the upper hemisphere ($z>0$) may be different
from that for the lower hemisphere ($z<0$). To make this point clear consider
for example the case of a dipole at the origin pointing in the $\hat z$
direction. Here $\Phi_{1,0}(x,y,z_{0})=-\Phi_{1,0}(x,y,-z_{0})$ so that the
expansion of $\Phi$ for the upper hemisphere differs in sign from the
expansion for the lower hemisphere. Therefore, the plane wave expansion for
$\Phi_{lm}$ should generally be written as%

\begin{equation}
\Phi_{lm}(x,y,z)=\frac1{4\pi^{2}}%
{\displaystyle\iint}
\tilde\Phi_{lm}^{\pm}(k_{x},k_{y})e^{i[k_{x}x+k_{y}y]}e^{-\sqrt{k_{x}%
^{2}+k_{y}^{2}}\left|  z\right|  }dk_{x}dk_{y} \label{eq1.2}%
\end{equation}
where the $+$ sign applies for $z>0$ and the $-$ sign for $z<0$. The
dependence on $z$ has been included by using $\nabla^{2}\Phi_{lm}=0$ and
demanding that $\lim\limits_{r\longrightarrow\infty}\Phi_{lm}=0$. A similar
expression for $q_{lm}$ reads:%

\begin{equation}
q_{lm}(x,y)=\frac{1}{4\pi^{2}}%
{\displaystyle\iint}
\tilde{q}_{lm}(k_{x},k_{y})e^{i[k_{x}x+k_{y}y]}dk_{x}dk_{y} \label{eq1.3}%
\end{equation}
When Eq.(\ref{eq1.2}) and Eq.(\ref{eq1.3}) are inserted into Eq.(\ref{eq1.1}),
and the identity
\[
\iint e^{i(k_{x}x+k_{y}y)}dxdy=4\pi^{2}\delta(k_{x})\delta(k_{x})
\]
is used, the result becomes
\begin{equation}
\Psi_{lm}(x,y,z)=\frac{1}{4\pi^{2}}%
{\displaystyle\iint}
\tilde{q}_{lm}(k_{x},k_{y})\tilde{\Phi}_{lm}^{\pm}(k_{x},k_{y})e^{i[k_{x}%
x+k_{y}y]}e^{-\sqrt{k_{x}^{2}+k_{y}^{2}}\left|  z\right|  }dk_{x}%
dk_{y}\text{.} \label{eq1.4}%
\end{equation}
In Appendix B we show that $\tilde{\Phi}_{lm}^{\pm}(k_{x},k_{y})$, as defined
by Eq.(\ref{eq1.2}), is given by:%

\begin{align}
\tilde{\Phi}_{lm}^{\pm}(k_{x},k_{y})  &  =\left\{
\begin{array}
[c]{cc}%
1 & ;z>0\\
(-1)^{l+m} & ;z<0
\end{array}
\right\}  \frac{2\pi}{i^{m}}\sqrt{\frac{4\pi}{2l+1}}\frac{(-1)^{m}}%
{\sqrt{(l+m)!(l-m)!}}\times\label{eq1.5}\\
&  \ (k_{x}^{2}+k_{y}^{2})^{(l-m-1)/2}(k_{x}+ik_{y})^{m}\text{.}\nonumber
\end{align}
Substituting Eq.(\ref{eq1.5}) for $\tilde{\Phi}_{lm}^{\pm}(k_{x},k_{y})$ into
Eq.(\ref{eq1.4}) gives
\begin{align}
\Psi_{lm}(x,y,z)  &  =\frac{1}{4\pi^{2}}\left\{
\begin{array}
[c]{cc}%
1 & ;z>0\\
(-1)^{l+m} & ;z<0
\end{array}
\right\}  \frac{2\pi}{i^{m}}\sqrt{\frac{4\pi}{2l+1}}\frac{(-1)^{m}}%
{\sqrt{(l+m)!(l-m)!}}\times\label{eq1.6}\\
&  \!%
{\displaystyle\iint}
(k_{x}^{2}+k_{y}^{2})^{(l-m-1)/2}(k_{x}+ik_{y})^{m}\tilde{q}_{lm}(k_{x}%
,k_{y})e^{i[k_{x}x+k_{y}y]}e^{-\sqrt{k_{x}^{2}+k_{y}^{2}}\left|  z\right|
}dk_{x}dk_{y}\nonumber
\end{align}

\subsection{The Discontinuity of the potential $\Psi$ across the Surface.}

The discontinuity of the potential across the surface (denoted by $\delta
\Psi_{lm}$) is defined as%

\begin{equation}
\delta\Psi_{lm}\equiv\Psi_{lm}(x,y,z=0^{+})-\Psi_{lm}(x,y,z=0^{-})\text{.}
\label{eq1.18}%
\end{equation}
Using Eq.(\ref{eq1.6}) in Eq.(\ref{eq1.18}) gives
\begin{align}
\delta\Psi_{lm}(x,y)  &  =\frac{2\pi}{i^{m}}\left[  1-(-1)^{l+m}\right]
\sqrt{\frac{4\pi}{2l+1}}\frac{(-1)^{m}}{\sqrt{(l+m)!(l-m)!}}\times
\label{eq1.19}\\
&  \left\{  \frac{1}{4\pi^{2}}%
{\displaystyle\iint}
(k_{x}^{2}+k_{y}^{2})^{(l-m-1)/2}(k_{x}+ik_{y})^{m}\tilde{q}_{lm}(k_{x}%
,k_{y})e^{i(k_{x}x+k_{y}y)}dk_{x}dk_{y}\right\} \nonumber\\
&  =\frac{2\pi}{i^{m}}\left[  1-(-1)^{l+m}\right]  \frac{1}{i^{m}}\sqrt
{\frac{4\pi}{2l+1}}\frac{(-1)^{m}}{\sqrt{(l+m)!(l-m)!}}\times\nonumber\\
&  \left[  \frac{\partial^{2}}{\partial x^{2}}+\frac{\partial^{2}}{\partial
y^{2}}\right]  ^{(l-m-1)/2}\left[  \frac{\partial}{\partial x}+i\frac
{\partial}{\partial y}\right]  ^{m}q_{lm}(x,y)\text{,}\nonumber
\end{align}
where in the last equality we have used the fact that a multiplication by
$i(k_{x}\pm ik_{y})$ in $(k_{x},k_{y})$-space is equivalent to the
differentiation $i\partial_{\pm}$ in $(x,y)$-space. Rearranging the
differentiation operators in Eq.(\ref{eq1.19}) yields
\begin{align*}
\delta\Psi_{lm}(x,y)  &  =\frac{2\pi}{i^{m}}\left[  1-(-1)^{l+m}\right]
\frac{1}{i^{m}}\sqrt{\frac{4\pi}{2l+1}}\frac{(-1)^{m}}{\sqrt{(l+m)!(l-m)!}%
}\times\\
&  (\partial_{+})^{\frac{1}{2}[(l-1)+m]}(\partial_{-})^{\frac{1}{2}%
[(l-1)-m]}q_{l,m}(x,y)\text{.}%
\end{align*}

Comparing this last result with Eq.(\ref{eq.6}) we first see that the
covariant form of Eq.(\ref{eq1.19}) is identical with what was predicted
earlier, and that the prefactor $b_{l,m}$ is given by
\begin{equation}
b_{l,m}=2\pi\left[  1-(-1)^{l+m}\right]  \sqrt{\frac{4\pi}{2l+1}}\frac
{1}{\sqrt{(l+m)!(l-m)!}}\text{.} \label{eq1.1.19}%
\end{equation}

For the case of $l=1,m=0$ for which $q_{1,0}(x,y)=\sqrt{3/4\pi}P_{z}$ (a
uniform double layer with density $P_{z}$) we find that $\delta\Psi_{lm}=4\pi
P_{z}$ as required.

\subsection{The Discontinuity of $E_{z}$ across the Surface.}

The discontinuity of the normal electric field across the surface (denoted by
$\delta(E_{z})_{l,m}$) is%

\begin{equation}
\delta(E_{z})_{l,m}\equiv\left[  -\frac{\partial}{\partial z}\Psi
_{lm}(x,y,z=0^{+})\right]  -\left[  -\frac{\partial}{\partial z}\Psi
_{lm}(x,y,z=0^{-})\right]  \label{eq1.20}%
\end{equation}
Thus, using Eq.(\ref{eq1.6}) we find that
\begin{align}
\delta(E_{z})_{l,m}(x,y)  &  =\frac{2\pi}{i^{m}}\left[  1+(-1)^{l+m}\right]
\sqrt{\frac{4\pi}{2l+1}}\frac{(-1)^{m}}{\sqrt{(l+m)!(l-m)!}}\times
\label{eq1.21}\\
&  \ \ \left\{  \frac{1}{4\pi^{2}}%
{\displaystyle\iint}
(k_{x}^{2}+k_{y}^{2})^{(l-m)/2}(k_{x}+ik_{y})^{m}\tilde{q}_{lm}(k_{x}%
,k_{y})e^{i(k_{x}x+k_{y}y)}dk_{x}dk_{y}\right\} \nonumber\\
\  &  =\frac{2\pi}{i^{m}}\left[  1+(-1)^{l+m}\right]  \frac{1}{i^{m}}%
\sqrt{\frac{4\pi}{2l+1}}\frac{(-1)^{m}}{\sqrt{(l+m)!(l-m)!}}\times\nonumber\\
&  \ \ \left[  \frac{\partial^{2}}{\partial x^{2}}+\frac{\partial^{2}%
}{\partial y^{2}}\right]  ^{(l-m)/2}\left[  \frac{\partial}{\partial x}%
+i\frac{\partial}{\partial y}\right]  ^{m}q_{lm}(x,y)\text{,}\nonumber
\end{align}
or
\begin{align}
\delta(E_{z})_{l,m}(x,y)  &  =2\pi\left[  1+(-1)^{l+m}\right]  \sqrt
{\frac{4\pi}{2l+1}}\frac{1}{\sqrt{(l+m)!(l-m)!}}\times\label{eq.1.1.21}\\
&  \ \ (\partial_{+})^{\frac{1}{2}(l+m)}(\partial_{-})^{\frac{1}{2}%
(l-m)}q_{lm}(x,y)\text{.}\nonumber
\end{align}

Comparing this result with Eq.(\ref{eq.5}) we see again that the covariant
form is identical with the one that was predicted earlier, and that the
prefactor $a_{l,m}$ is
\begin{equation}
a_{l,m}=2\pi\left[  1+(-1)^{l+m}\right]  \sqrt{\frac{4\pi}{2l+1}}\frac
{1}{\sqrt{(l+m)!(l-m)!}}\text{.} \label{1.2.21}%
\end{equation}
Note that $a_{l,m}$ vanishes for odd ($l+m$) whereas $b_{l,m}$ vanishes for
even ($l+m$). Consequently, for a given ($l,m$) there can be only
discontinuity either in the normal field or in the potential but not in both simultaneously.

\section{Summary}

In this paper we derived expressions for the electrostatic field discontinuity
across a flat surface in vacuum, having general ($l^{th},m^{th}$) multipole
surface charge density. We showed that dimensional analysis and the principle
of covariance alone are suffice to determine the form of the expression. The
exact expressions for the field discontinuity were then derived mathematically
and were shown to confirm with our early prediction. These expressions are
\begin{align}
\delta E_{z}(x,y)  &  =%
{\displaystyle\sum\limits_{l=0}^{\infty}}
{\displaystyle\sum\limits_{m=-l}^{l}}
2\pi\left[  1+(-1)^{l+m}\right]  \sqrt{\frac{4\pi}{2l+1}}\frac{1}%
{\sqrt{(l+m)!(l-m)!}}\times\label{eqs.1}\\
&  \ \ (\partial_{+})^{\frac{1}{2}(l+m)}(\partial_{-})^{\frac{1}{2}%
(l-m)}q_{l,m}(x,y)\text{,}\nonumber
\end{align}
for the discontinuity of the normal electric field, and
\begin{align*}
\delta\Psi(x,y)  &  =%
{\displaystyle\sum\limits_{l=0}^{\infty}}
{\displaystyle\sum\limits_{m=-l}^{l}}
2\pi\left[  1-(-1)^{l+m}\right]  \sqrt{\frac{4\pi}{2l+1}}\frac{1}%
{\sqrt{(l+m)!(l-m)!}}\times\\
&  \ \ (\partial_{+})^{\frac{1}{2}[(l-1)+m]}(\partial_{-})^{\frac{1}%
{2}[(l-1)-m]}q_{l,m}(x,y)\text{,}%
\end{align*}
for the discontinuity of the potential. Here, $q_{l,m}(x,y)$ is the
$(l^{th},m^{th})$ multipolar surface charge density over the plane,
$\partial_{\pm}\equiv\partial/\partial x\pm i\partial/\partial y$, $E_{z}$ is
the normal electric field and $\Psi$ is the potential.

It is interesting to note how these expressions can be extended to the
\emph{electrodynamic} case (especially to the discontinuity of the magnetic
field). To give the reader a sense of the subtleties that arise when
considering the electrodynamic case we point out that, here, further
complication arises as there is an additional parameter that must be taken
into account, namely $k_{0}$, the wave number. The arguments that led us
earlier to a unique form for the discontinuity relation are no longer valid.
Instead, we must now incorporate the presence of $k_{0}$ into the expressions
in a covariant form. This procedure, however, is not uniquely determined as is
shown by the following argument: The operator $(\partial_{+})^{N}$ must now be
replaced by an operator that has similar dimension, transforms as
$(\partial_{+})^{N}$ and may include $k_{0}$ as an additional parameter. The
general operator that obeys these restrictions is $P_{N}(\partial_{+}%
,k_{0}e^{-i\varphi})$ where $P_{N}(x,y)$ is a general two-dimensional
polynomial of degree $N$ and $\varphi$ is $Tan^{-1}(y/x)$. Similarly,
$(\partial_{-})^{N}$ should be replaced by $Q_{N}(\partial_{-},k_{0}%
e^{+i\varphi})$ where $Q_{N}(x,y)$ is yet another two dimensional polynomial.
Covariance and dimensional analysis alone cannot yield the coefficients of
$P_{N}$ and $Q_{N}$.

Yet another interesting question is how does the discontinuity of the fields
depends (if any) on the curvature of the surface. Recall that the
discontinuity in the normal electric field due to a surface charge density, as
well as the discontinuity in the potential due to a surface dipolar density
\emph{does not} depend on the curvature of the surface. One might then
carelessly conclude that for higher multipolar densities this is the case too.
However, the existence of the differential operators in Eq.(\ref{eqs.1}),
which do not appear at $(l,m)=(0,0)$ and $(1,0)$, might suggest that for
higher multiples the curvature \emph{does} appear. We conjecture that the
curvature \emph{does} enter in the form of a covariant differentiation which
replaces the $\partial_{\pm}$ operators.

\pagebreak 

\appendix

\section{Transformation rule for $q_{l,m}(x,y)$ and $\partial_{\pm}$.}

When the system of coordinate is rotated by an angle $\alpha$ around the
$z$-axis, the coordinates transform as
\begin{equation}%
\begin{array}
[c]{c}%
r\rightarrow r^{\prime}=r\\
\varphi\rightarrow\varphi^{\prime}=\varphi-\alpha\\
\partial/\partial r\rightarrow\partial/\partial r^{\prime}=\partial/\partial
r\\
\partial/\partial\varphi^{\prime}=\partial/\partial\varphi
\end{array}
\label{Eq.A.2}%
\end{equation}
Thus, while the potential due to a multipolar charge $q_{l,m}$ at the origin
is
\[
\Phi_{l,m}(r,\theta,\varphi)=q_{l,m}\frac{4\pi}{2l+1}\frac{Y_{l,m}%
(\theta,\varphi)}{r^{l+1}}%
\]
as expressed with the old coordinate system, the potential in the new
coordinate system is
\[
\Phi_{l,m}^{\prime}(r^{\prime},\theta^{\prime},\varphi^{\prime})=q_{l,m}%
^{\prime}\frac{4\pi}{2l+1}\frac{Y_{l,m}(\theta^{\prime},\varphi^{\prime}%
)}{r^{\prime l+1}}%
\]
Since they must give the same value we have
\[
q_{l,m}Y_{l,m}(\theta,\varphi)=q_{l,m}^{\prime}Y_{l,m}(\theta^{\prime}%
,\varphi^{\prime})=q_{l,m}^{\prime}Y_{l,m}(\theta,\varphi-\alpha)
\]
Using the properties of $Y_{l,m}()$ which depends on $\varphi$ only through
$e^{im\varphi}$ we have
\[
q_{l,m}Y_{l,m}(\theta,\varphi)=q_{l,m}^{\prime}Y_{l,m}(\theta,\varphi
)e^{-im\alpha}%
\]
which shows that $q_{l,m}^{\prime}$ must be given by
\[
q_{l,m}^{\prime}=q_{l,m}e^{im\alpha}%
\]
We now turn to find the transformation rule for $\partial_{\pm}$. Rewriting
$\partial_{\pm}$ in terms of polar coordinate system yields
\begin{equation}
\partial_{\pm}\equiv\frac{\partial}{\partial x}\pm i\frac{\partial}{\partial
y}=e^{\pm i\varphi}\left[  \frac{\partial}{\partial r}\pm\frac{i}{r}%
\frac{\partial}{\partial\varphi}\right]  \label{eq.A.3}%
\end{equation}
When these are used with Eq.(\ref{Eq.A.2}) one finds that
\[
\partial_{\pm}^{\prime}=e^{\pm i\varphi^{\prime}}e^{\mp i\alpha}\left[
\frac{\partial}{\partial r}\pm\frac{i}{r}\frac{\partial}{\partial\varphi
}\right]  =e^{\mp i\alpha}\partial_{\pm}%
\]

\section{Derivation of $\tilde\Phi_{l,m}^{\pm}(k_{x},k_{y})$.}

In this section we show that $\tilde{\Phi}_{lm}^{\pm}(k_{x},k_{y})$ as given
by Eq.(\ref{eq1.6}) is indeed the Fourier transform of $\Phi_{lm}(x,y,z=0)$.
The Fourier transform of $\Phi_{lm}$ over the $x$-$y$ plane may be calculated
straighforwardly by using the following definite integral\cite{grad}
\[%
\begin{array}
[c]{c}%
{\displaystyle\int\limits_{0}^{\infty}}
x(a^{2}+x^{2})^{-\frac{1}{2}\mu}P_{\mu-1}^{-\nu}\left[  \frac{a}{\sqrt
{a^{2}+x^{2}}}\right]  J_{\nu}(xy)dx=\frac{y^{\mu-2}e^{-ay}}{\Gamma(\mu+\nu
)}\text{ }\\
\text{(}Re(a)>0\text{, }y>0\text{, }Re(\nu)>-1\text{, }Re(\mu)>\frac{1}%
{2}\text{)}.
\end{array}
\]
However we prefer to show it by explicitly evaluating the inverse Fourier transform.

Substitution of Eq.(\ref{eq1.6}) inside Eq.(\ref{eq1.2}) gives:%

\begin{align}
I  &  \equiv\frac{1}{4\pi^{2}}%
{\displaystyle\iint}
\tilde{\Phi}_{lm}^{\pm}(k_{x},k_{y})e^{i[k_{x}x+k_{y}y]}e^{-\sqrt{k_{x}%
^{2}+k_{y}^{2}}\left|  z\right|  }dk_{x}dk_{y}\label{eq.C.1}\\
\  &  =\left\{
\begin{array}
[c]{cc}%
1 & ;z>0\\
(-1)^{l+m} & ;z<0
\end{array}
\right\}  \frac{1}{2\pi i^{m}}\sqrt{\frac{4\pi}{2l+1}}\frac{(-1)^{m}}%
{\sqrt{(l+m)!(l-m)!}}\times\nonumber\\
&  \
{\displaystyle\iint}
(k_{x}^{2}+k_{y}^{2})^{(l-m-1)/2}(k_{x}+ik_{y})^{m}e^{i[k_{x}x+k_{y}%
y]}e^{-\sqrt{k_{x}^{2}+k_{y}^{2}}\left|  z\right|  }dk_{x}dk_{y}%
\text{.}\nonumber
\end{align}
By using polar coordinates (defined by $k_{r}\equiv\sqrt{k_{x}^{2}+k_{y}^{2}}$
and $\varphi_{k}\equiv Tg^{-1}(k_{y}/k_{x})$) Eq.(\ref{eq.C.1}) becomes:
\begin{align}
I  &  =\left\{
\begin{array}
[c]{cc}%
1 & ;z>0\\
(-1)^{l+m} & ;z<0
\end{array}
\right\}  \frac{1}{2\pi i^{m}}\sqrt{\frac{4\pi}{2l+1}}\frac{(-1)^{m}}%
{\sqrt{(l+m)!(l-m)!}}\times\label{eq.C.2}\\
&  \ \ \int\limits_{0}^{\infty}k_{r}dk_{r}\int\limits_{0}^{2\pi}d\varphi
_{k}k_{r}^{(l-1)}e^{im\varphi_{k}}e^{ik_{r}r\sin\theta\cos(\varphi_{k}%
-\varphi)}e^{-k_{r}r\left|  \cos\theta\right|  }\nonumber
\end{align}
The integral over the angular part is easily evaluated by using\cite{jack2}%

\begin{equation}
J_{m}(x)=\frac{1}{2\pi i^{m}}\int\limits_{0}^{2\pi}e^{ix\cos\varphi-im\varphi
}d\varphi\text{.} \label{eq.C.3}%
\end{equation}
This gives:%

\begin{align}
I  &  =\left\{
\begin{array}
[c]{cc}%
1 & ;z>0\\
(-1)^{l+m} & ;z<0
\end{array}
\right\}  \frac{1}{2\pi i^{m}}\sqrt{\frac{4\pi}{2l+1}}\frac{(-1)^{m}}%
{\sqrt{(l+m)!(l-m)!}}\times\label{eq.C.4}\\
&  \ \ \int\limits_{0}^{\infty}k_{r}dk_{r}k_{r}^{(l-1)}e^{im\varphi}2\pi
i^{m}J_{m}(k_{r}r\sin\theta)e^{-k_{r}r\left|  \cos\theta\right|  }\nonumber
\end{align}
Since\cite{grad}
\begin{align}
\int\limits_{0}^{\infty}x^{\mu-1}J_{\nu}(\beta x)e^{-\alpha x}dx  &
=(\alpha^{2}+\beta^{2})^{-\mu/2}\Gamma(\mu+\nu)P_{\mu-1}^{-\nu}(\alpha
/\sqrt{\alpha^{2}+\beta^{2}})\label{eq.C.5}\\
\alpha,\beta &  >0\text{ };Re(\mu+\nu)>0\text{,}\nonumber
\end{align}
we get
\begin{align}
I  &  =\left\{
\begin{array}
[c]{cc}%
1 & ;z>0\\
(-1)^{l+m} & ;z<0
\end{array}
\right\}  \frac{1}{2\pi i^{m}}\sqrt{\frac{4\pi}{2l+1}}\frac{(-1)^{m}}%
{\sqrt{(l+m)!(l-m)!}}\times\label{eq.C.6}\\
&  \ \ e^{im\varphi}2\pi i^{m}r^{-(l+1)}\Gamma(l+m+1)P_{l}^{-m}(\left|
\cos\theta\right|  )\nonumber
\end{align}
By using \cite{jack2}
\begin{equation}
P_{l}^{-m}(x)=(-1)^{m}\frac{(l-m)!}{(l+m)!}P_{l}^{m}(x) \label{eq.C.7}%
\end{equation}
and the definition of $\Gamma(x)$ by which $\Gamma(l+m+1)=(l+m)!$ we find
that
\begin{align}
I  &  =\left\{
\begin{array}
[c]{cc}%
1 & ;z>0\\
(-1)^{l+m} & ;z<0
\end{array}
\right\}  \frac{1}{2\pi i^{m}}\sqrt{\frac{4\pi}{2l+1}}\frac{(-1)^{m}}%
{\sqrt{(l+m)!(l-m)!}}\times\label{eq.C.8}\\
&  \ \ e^{im\varphi}2\pi i^{m}r^{-(l+1)}(l+m)!(-1)^{m}\frac{(l-m)!}%
{(l+m)!}P_{l}^{m}(\left|  \cos\theta\right|  )\nonumber\\
\  &  =e^{im\varphi}\sqrt{\frac{4\pi}{2l+1}}r^{-(l+1)}\sqrt{\frac
{(l-m)!}{(l+m)!}}\left\{
\begin{array}
[c]{c}%
1\\
(-1)^{l+m}%
\end{array}
\right\}  P_{l}^{m}(\left|  \cos\theta\right|  )\nonumber
\end{align}
Using the properties of $P_{l}^{m}(x)$ under inversion it may be easily
verified that%

\begin{equation}
\left\{
\begin{array}
[c]{cc}%
1 & ;z>0\\
(-1)^{l+m} & ;z<0
\end{array}
\right\}  P_{l}^{m}(\left|  \cos\theta\right|  )=P_{l}^{m}(\cos\theta)\text{,}
\label{eq.C.9}%
\end{equation}
and since \cite{jack2}
\begin{equation}
Y_{lm}(\theta,\varphi)=\sqrt{\frac{2l+1}{4\pi}\frac{(l-m)!}{(l+m)!}}P_{l}%
^{m}(\cos\theta)e^{im\varphi}\text{,} \label{eq.C.10}%
\end{equation}
we finally find, as required, that
\begin{equation}
I=\frac{4\pi}{2l+1}\frac{Y_{lm}(\theta,\varphi)}{r^{l+1}}=\Phi_{lm}(\vec{r})
\label{eq.C.11}%
\end{equation}

\end{document}